\documentclass[a4paper,12pt,twoside]{article}
\usepackage{amsmath}
\usepackage{amssymb}
\usepackage[dvips]{graphicx}
\begin{document}
\author{P. Konrad, H. Lenske\\
Institut f\"ur Theoretische Physik, Universit\"at Gie\ss en\\
Heinrich-Buff-Ring 16, D-35392 Gie\ss en, Germany}

\title{Correlations in hypernuclear matter}
\maketitle

\begin{abstract}
We investigate short range correlations in nuclear and hypernuclear
matter. Self-energies due to short range correlations and their
influence on the nucleon and $\Lambda$-hyperon spectral functions
are described in an approach accounting for a realistic treatment of
mean-field dynamics and a self-consistently derived quasi-particle
interaction. Landau-Migdal theory is used to derived the short range
interaction from a phenomenological Skyrme energy density
functional, subtracting the long range pionic contributions to the
nucleonic spectral functions. We discuss our results for different
hyperon-baryon ratios to show the influence of strangeness on the
correlations in hypernuclear matter.
\end{abstract}

\section{Introduction}
Dynamical correlations going beyond static mean-field calculations are of interest for understanding of the dynamics in a nuclear many-body system. For nuclear matter this has been discussed extensively in literature ~\cite{Dickhoff:2004xx,Lehr:2001qy,Froemel:2003dv,Konrad:2005qm}. In the case of hypernuclear matter the lack of knowledge of the hyperon interaction make it very difficult to investgate dynamical correlations. In our previous work we extracted information on the particle-hole interaction from the mean-field leading to an effective quasiparticle interaction. Therefore, we are able to extend our approach into the strangeness sector. 
In section 2 be summarize the basic equation for the self-energies and the spectral functions. It is also shown how to extend the approach to the case of hypernuclear matter. 
Some results for the on-shell widths and the spectral functions are shown and discussed for various strangeness fractions in section 3.
Finally we will close in section 4 with a summary.

\section{The model}
\subsection{The spectral functions and the self-energy in hypernuclear matter}
Correlations in hypernuclear matter are described by a transport-theoretical approach.
The model was presented in \cite{Lehr:2001qy} and used for calculations in nuclear matter ~\cite{Lehr:2001qy,Froemel:2003dv,Konrad:2005qm}. We restrict ourselves to a short summary, focusing on the changes with resepct to hypernuclear matter consisting of prontons, neutrons and $\Lambda$-hyperons. A detailed discussion of our model can be found in \cite{Lehr:2001qy} and the underlying Green's function formalism is described in much detail in \cite{Kadanoff:1962st}. 
In contrast to normal nuclear matter in hypernuclear matter one has to take into account contributions from nucleons and $\Lambda$-hyperons to the polarization self-energies. Therefore two-particle-one-hole (2p1h) and the one-particle-two-hole (1p2h) polarization self-energies $\Sigma^>$ and $\Sigma^<$ for $q=n,p,\Lambda$ in hypernuclear matter can be expressed in terms of dynamical self-energies:
	\begin{eqnarray}
	\Sigma^{\gtrless}_q
	=&&\Sigma^{\gtrless}_{qn\bar{n}}+\Sigma^{\gtrless}_{qp\bar{p}}+\Sigma^{\gtrless}_{q\Lambda\bar{\Lambda}}\\
	=&&g \sum_{q'=n,p,\Lambda}\int \frac{d^3k_2 d\omega_2}{(2\pi)^4}
    	\frac{d^3k_3 d\omega_3}{(2\pi)^4}\frac{d^3k_4 d\omega_4}{(2\pi)^4}
    	(2\pi)^4\delta^4(k+k_2-k_3-k_4)\overline{|M_{qq'}|^2}\nonumber\\
    	&&\times g^{\lessgtr}_{q'}(\omega_2,k_2)g^{\gtrless}_{q}(\omega_3,k_3)
    	g^{\gtrless}_{q'}(\omega_4,k_4)\nonumber
	\label{eq:sigma}
	\end{eqnarray}
where $g=2$ is the spin degeneracy and $q'$ runs over protons, neutrons and $\Lambda$-hyperons. The delta-function conserves momentum and energy. The matrix element $\overline{|M_{qq'}|^2}$ is the square of the baryon scattering amplitude averaged over  the spin of the incoming and outgoing particles. We assumed it to be independent of energy and momentum, accounting for short-range effects, since a constant factor in phase space corresponds to a delta-function in coordinate space. As we showed before this assumption gives good results for nuclear matter \cite{Lehr:2001qy,Froemel:2003dv,Konrad:2005qm} . We leave the structure of $\overline{|M_{qq'}|^2}$ open for the moment, but address it later.

The nonrelativistic spectral function is found explicity as
 \begin{eqnarray}
 a_q(\omega,k)=\frac{\Gamma_q(\omega,k)}
            {(\omega-\frac{k^2}{2 m_q}-\Sigma^{mf}_q-\mathrm{Re}
            \Sigma^{ret}_q(\omega,k))^2+\frac{1}{4}\Gamma^2_q(\omega,k)},
\label{eq:spectral} 
	\end{eqnarray} where $\mathrm{Re}\Sigma^{ret}_q(\omega,k)$
is the real part of the retarded self-energy, which is calculated by a dispersion relation.

Finally the width $\Gamma_q(\omega,k)$ is given by the imaginary part of the
retarded self-energy, 
	\begin{eqnarray}
    	\Gamma_q(\omega,k)=2 \mathrm{Im}\Sigma^{ret}_{q}=
    	i(\Sigma^>_q(\omega,k)-\Sigma^<_q(\omega,k)).
	\end{eqnarray}

We obtain the mean-field from a Skyrme energy density functional. For our calculations we use the parameter set SLy230a from \cite{Chabanat:1997qh} for nucleons and the parameter sets Ly-IV and SLL1 from \cite{Mornas:2004vs} for the $N$-$\Lambda$ and $\Lambda$-$\Lambda$ interaction respectively.
A detailed dicussion of the Sykrme interaction of hyperons can be found in \cite{Mornas:2004vs}. 

In the numerical calculations we use an iterative approach. Starting with an initial choice for the widths, i.e. the imaginary part of the self-energy, we calculate the spectral functions, which then serve as input for recalculating the self-energies. The calculation is stopped, when the spectral functions are converged to a given accuracy. In the language of Feynman diagrams this approach corresponds to the summation of the sunset diagram to all orders ~\cite{Lehr:2001qy}.

\subsection{The short range interaction}
An appropriate way to determine the structure of the average matrix elements $\overline{|M_{qq'}|^2}$ is provided by Landau-Migdal Theory.
The N-N interaction is represented by a superposition of the spin-isospin operators 
\begin{eqnarray}
F_{NN'}(\rho)=f_{NN'}(\rho)+f'_{NN'}(\rho)\vec{\tau_1}\vec{\tau_2}+g_{NN'}(\rho)\vec{\sigma_1}\vec{\sigma_2}+g'_{NN'}(\rho)\vec{\tau_1}\vec{\tau_2}\vec{\sigma_1}\vec{\sigma_2}
\end{eqnarray}
and for the $\Lambda$-N and $\Lambda-\Lambda$ interaction, respectively,
\begin{eqnarray}
F_{N\Lambda}(\rho)=f_{N\Lambda}(\rho)+g_{N\Lambda}(\rho)\vec{\sigma_1}\vec{\sigma_2}\\
F_{\Lambda\Lambda}(\rho)=f_{\Lambda\Lambda}(\rho)+g_{\Lambda\Lambda}(\rho)\vec{\sigma_1}\vec{\sigma_2}
\end{eqnarray}
Here, only isoscalar interaction contribute because the $\Lambda$-hyperon is an isospin singelt state.
We subtracted the pion part from from the N-N interaction, thus removing the long-range interactions \cite{Konrad:2005qm}. The remaining interaction is then determined by short-range effects only.
The average matrix element $\overline{|M_{qq'}|}$ we define by wieghting with the spin-multiplicities
\begin{eqnarray}
	\overline{|M_{qq'}|}=\sqrt{\frac{1}{4}(f_{qq'}^2+3g_{qq'}^2)},
\end{eqnarray}
where $q$ and $q'$ are running over nucleons and  $\Lambda$-hyperons.

\begin{figure}
  \begin{center}
    \begin{tabular}{cc}
      \resizebox{60mm}{!}{\includegraphics{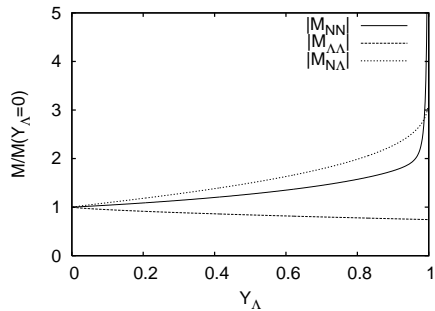}} &
    \end{tabular}
    \caption{The average matrix elements depending on the strangeness fraction. The matrix elements are normalized to pure nuclear matter. }
    \label{hypermeq}
  \end{center}
\end{figure}

Figure \ref{hypermeq} shows the nucleon-nucleon, $\Lambda$-nucleon and $\Lambda$-$\Lambda$ average matrix element $\overline{|M_{qq'}|}$ in dependence on the strangeness fraction $Y_{\Lambda}=\frac{\rho_{\Lambda}}{\rho_B}$, where $\rho_B=\rho_p+\rho_n+\rho_{\Lambda}$ is the local baryon number density. Reduced Pauli-blocking leads to in increase of the nucleon-nucleon and $\Lambda$-nucleon matrix element, since with increasing strangeness less nucleon states are occupied. For the $\Lambda$-$\Lambda$ matrix element the picture looks different: with increasing strangeness fraction more $\Lambda$ states are occupied, this leads to an decrease of  $\Lambda$-$\Lambda$ interactions. These results are in agreement with results for the nucleon-nucleon interaction obtained in nuclear matter \cite{Konrad:2005qm} 

\section{Results}
To investigate the influence of correlations in strange matter  we compare results at low and a high strangeness fraction by keeping the number density fixed. Here we compare results for strangeness fraction $Y_{\Lambda}=0.1$ and $Y_{\Lambda}=0.5$, while the total density is fixed to nuclear matter density $\rho_B=0.16 fm^{-3}$.  We note that we do not distinguish between protons and neutrons.

\begin{figure}
  \begin{center}
    \begin{tabular}{cc}
      \resizebox{60mm}{!}{\includegraphics{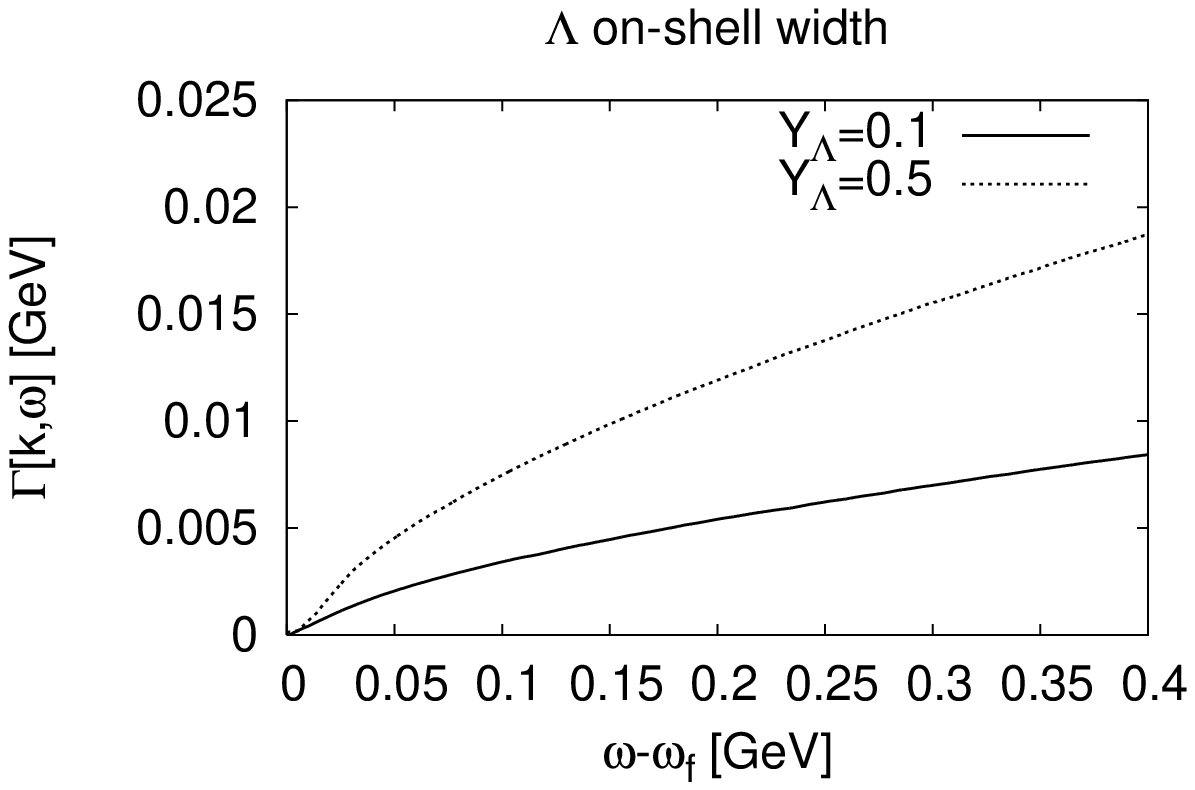}} &
      \resizebox{60mm}{!}{\includegraphics{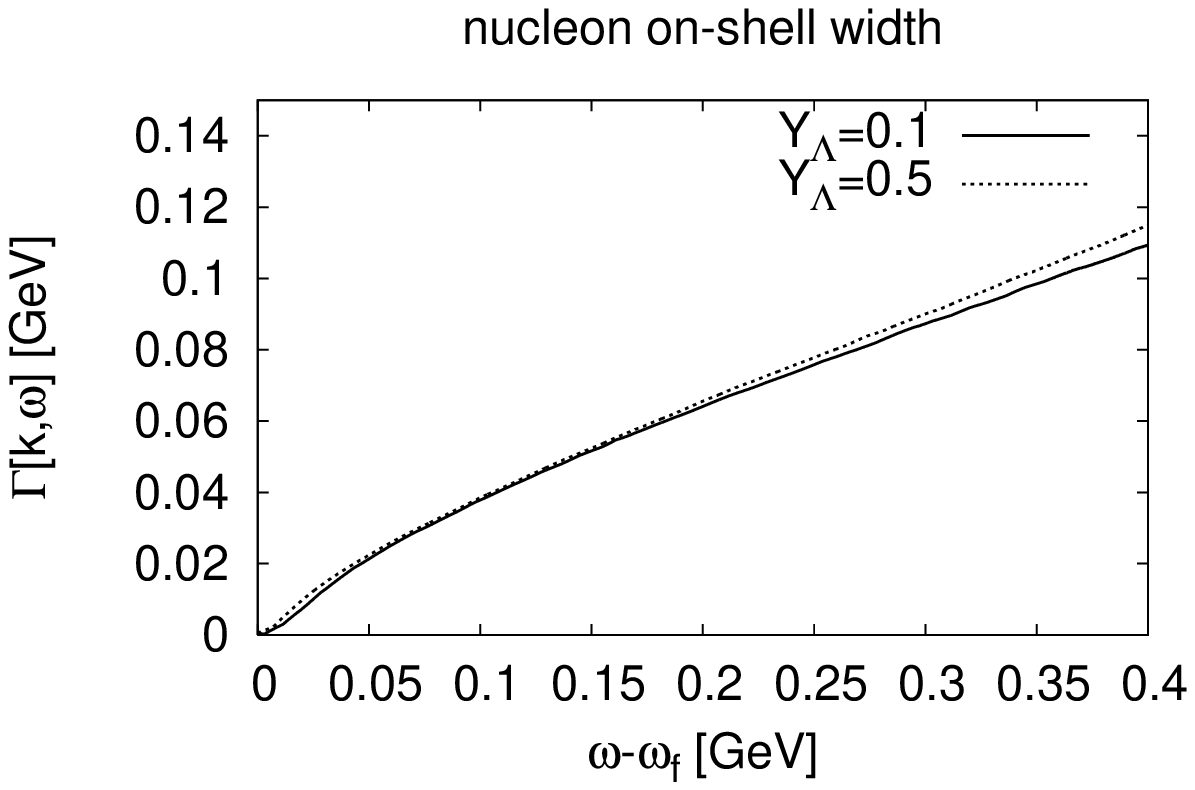}} \\
    \end{tabular}
    \caption{On-shell width of nucleons and $\Lambda$-hyperons for different strangeness fractions.}
    \label{onshell}
  \end{center}
\end{figure}

Dynamical correlations are reflected most clearly and directly in the imaginary part of the self-energies. Figure \ref{onshell} shows the imaginary part of self-energies evaluated at the on-shell point $\omega=\omega(k)$, defining the on-shell width.
The global energy dependence shows a strong increase of  the on-shell width at energies away from the Fermi surface. The fact that the self-energy vanishes at the Fermi surface reflects the stability of the ground state. On a quantitative level there is a big difference between the nucleons and $\Lambda$-hyperons. Figure \ref{onshell} shows a much smaller on-shell width for $\Lambda$-hyperons than for nucleons. This fact can be explained by the much smaller average matrix element of the hyperons, reducing considerably the coupling to the excitations of the medium.

Comparing the results for different strangeness fraction shows only a slightly increase of the on-shell widths of a few MeV in case of higher strangeness fraction. This increase can be explained by the enhanced contributions from nucleon-nucleon and $\Lambda$-nucleon interaction, so that correlations with nucleons become strong. But in an overall view strangeness seems to influence the correlations only at a very low level. This leads to the conclusion that hypernuclear matter can be seperated into nucleon and $\Lambda$-hyperon subsystems which are almost decoupled.

\begin{figure}
  \begin{center}
    \begin{tabular}{cc}
      \resizebox{60mm}{!}{\includegraphics{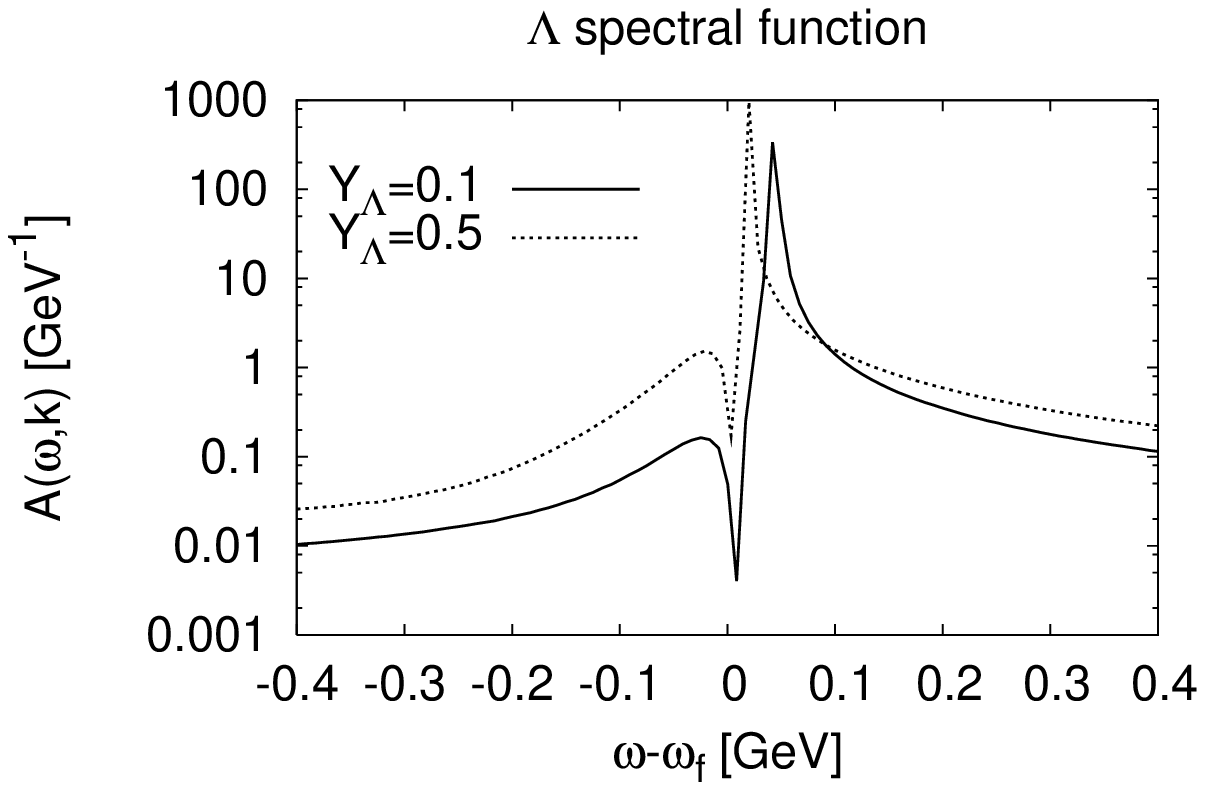}} &
      \resizebox{60mm}{!}{\includegraphics{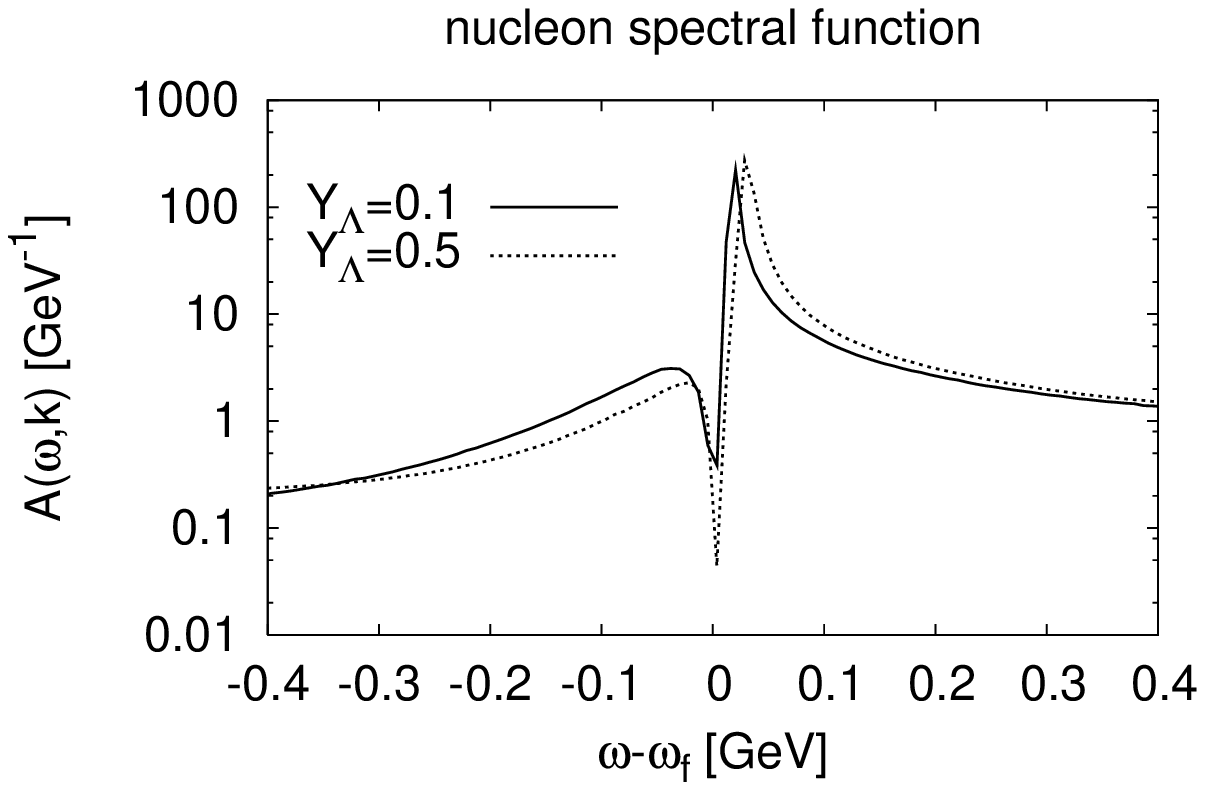}} \\
    \end{tabular}
    \caption{Spectral functions of nucleons and $\Lambda$-hyperons for different strangeness fractions at $k=312.5$ MeV/c .}
    \label{spectral}
  \end{center}
\end{figure}

In figure \ref{spectral} the spectral functions of nucleons and $\Lambda$-hyperons are displayed at momentum $k=312.5$ Mev/c which is slightly above the Fermi-momentum. One clearly sees the quasiparticle peaks above the Fermi surface. Because of the weaker hyperon interaction the spectral function of the $\Lambda$-hyperons is more concentrated at the quasi-particle pole than for the nucleons. A slight shifting of the quasiparticle peaks is found for different strangeness ratios. This can be explained by a change of the mean-field through the hyperons.

\section{Summary}
Dynamical correlations in hypernuclear matter were investigated in a self consistent approach, extending our previous work into a new regime. It is worthwhile to emphasize that this is the first systematic investgation of dynamical correlations in hypernuclear matter. 

Mean-field effects were included by  using a  modern Skyrme energy density functional with parameters adjusted both to finite hypernuclei and hypernuclear matter. Using Landau-Migdal theory we derived from the Skyrme functional the corresponding effective quasiparticle interaction. 

We compared the results for the on-shell widths and for the spectral functions for different strangeness ratios. The overall features of the self-energies and the spectral function resemble those found in pure nuclear matter. Our results lead to the conclusion that correlations via nucleons play the most important role, implying a separation of hypermatter in nucleon and $\Lambda$-hyperon subsystems which are only weakly interacting. This observation agrees perfectly well with the sharp spectral structures seen in hypernuclear spectra, e. g. \cite{Hotchi:2001rx}.

\end{document}